\documentclass[preprint]{elsarticle}
\usepackage{epstopdf}
\usepackage{caption}
\usepackage{setspace}
\usepackage{amssymb}
\usepackage{multirow}
\usepackage{amsthm}
\usepackage{booktabs}
\usepackage{amsmath}
\usepackage{amssymb}
\usepackage{mathrsfs}
\usepackage{longtable}
\usepackage{lscape}
\usepackage{url}
\graphicspath{{figs/}{figsgaoerb/}} 
\usepackage{tabularx}
\usepackage{epsfig}
\usepackage{colortbl}
\usepackage{subfigure}
\usepackage{tikz,mathpazo}
\usepackage{pgfplots}
\usepackage{graphicx}
\usepackage{natbib}
\usepackage[noend]{algpseudocode}
\usepackage{algorithm}
\usepackage{algorithmicx}   
\usepackage{array}  
\usepackage{hyperref}

\usetikzlibrary{shapes.geometric, arrows}
\newtheorem{mydef}{Definition}
\begin{document}
	\begin{frontmatter}
		\title{A novel weighted approach for time series forecasting based on visibility graph}
	
	\author[address1]{Tianxiang Zhan}
	\author[address1,address2,address3]{Fuyuan Xiao \corref{mycorrespondingauthor}}
	
	\address[address1]{School of Computer and Information Science, Southwest University, Chongqing, 400715, China}
	\address[address2]{School of Big Data and Software Engineering, Chongqing University, Chongqing, 401331, China}
	\address[address3]{National Engineering Laboratory for Integrated Aero-Space-Ground-Ocean Big Data Application Technology, China}
	\cortext[mycorrespondingauthor]{Corresponding author: Fuyuan Xiao is with the School of Big Data and Software Engineering, Chongqing University, Chongqing 401331, China. (e-mail: xiaofuyuan@cqu.edu.cn)
	}
		\begin{abstract}
		Time series has attracted a lot of attention in many fields today. Time series forecasting algorithm based on complex network analysis is a research hotspot. How to use time series information to achieve more accurate forecasting is a problem. To solve this problem, this paper proposes a weighted network forecasting method to improve the forecasting accuracy. Firstly, the time series will be transformed into a complex network, and the similarity between nodes will be found. Then, the similarity will be used as a weight to make weighted forecasting on the predicted values produced by different nodes. Compared with the previous method, the proposed method is more accurate. In order to verify the effect of the proposed method, the experimental part is tested on M1, M3 datasets and Construction Cost Index (CCI) dataset, which shows that the proposed method has more accurate forecasting performance.
		\end{abstract}
	\end{frontmatter}

	\section{Introduction}
	
	Time series is a common data format, and there are many typical data, such as wind speed, stock price, seismic waves and so on. Time series forecasting methods can predict a value at a future node in time. And time series forecasting can also provide reference value for economic evaluation \cite{huang2021natural, hewamalage2022global, huang2021new, cheng2022financial}, computation optimizer \cite{cheng2021high,bandara2021improving, le2022deep}, cost forecasting \cite{huang2021new2, DBLP:journals/ijis/ZhanX21} and so on \cite{DBLP:journals/apin/SongX22, ilic2021explainable, chen2021bayesian}.
	
	Complex networks are the hotspot of recent research and have a wide range of influences. Challenging problems in complex network research include, but are not limited to, influence communicator identification \cite{DBLP:journals/isci/WenD20, DBLP:journals/isci/LiX21, DBLP:journals/isci/ShangDC21, DBLP:journals/tfs/ZhangZDC22}, fractal analysis \cite{DBLP:journals/inffus/WenC21, zhang2022betweenness}, decision making \cite{ribas2020fusion,DBLP:journals/isci/LiuHD21}, network evolution \cite{DBLP:journals/tec/WangMWXKC21,9852288, DBLP:journals/kbs/YangX21, DBLP:journals/eaai/ChenDC21}, quantum theory \cite{DBLP:journals/tnn/Xiao21, DBLP:journals/tcyb/Xiao22} ,network clustering \cite{li2021multivariate} and so on \cite{9721052, shang2022effective}. Information inherited from the network can help us better understand the properties of time series and make better forecastings. Therefore, a lot of research has focused on how to model time series in the form of networks. The visibility graph (VG) algorithm \cite{lacasa2008time}, proposed by Lacasa et al., converts time series into complex networks. Visibility graph have a wide range of applications such as image processing, decision-making and so on. In the study of time series  forecasting, Zhang et al. proposed the forecasting method \cite{zhang2017novel} using VG and the superposed random walk (SRW) algorithm \cite{liu2010link} . Mao and Xiao improved the forecasting method of Zhang et al \cite{mao2019time}.
	
	The previous method of Zhang et al. \cite{zhang2017novel} and the method of Mao and Xiao \cite{mao2019time} only considered the node most similar to the last node of the time series in the complex network when forecasting the time series. Because the structure of time series is variable, considering only one node for the final forecast will reduce the accuracy of the forecast. The previous research \cite{zhang2017novel} uses the similarity of nodes as a reference, so the proposed method uses the similarity as a weight to weight the predicted values of different nodes, and obtains a more accurate forecasting effect than the previous method.
	
	In this paper, there are some fundamental theories in Section 2. In Section 3, the paper provide the explanation of the proposed method. In Section 4, experimental results and analysis are introduced. In Section 5, it draws a conclusion of the paper.

	\section{Preliminaries}
	This section will introduce some preliminaries, including VG and link forecasting.

	\subsection{Visibility graph}
	A given time series U is as follows: 
	\begin{equation}
		U=\left \{ \left ( t_1,y_1 \right ),\left ( t_2,y_2 \right ),...,\left ( t_m,y_m \right ),...,\left ( t_n,y_n \right ) \right \}
	\end{equation}
	where $y_m$ is the actual value of the data, and $t_m$ represents the node in time. In the visibility graph, an element $\left ( t_m,  y_m \right )$ is defined as a node.
	
	 In a complex network model, the relationship between nodes needs to be considered. So the nodes are analyzed in pairs. If two nodes $\left ( t_a, y_a \right )$ and $\left ( t_b, y_b \right )$ are connected with an edge, any node between them $\left ( t_c, y_c \right )$ is required to satisfy  \cite{lacasa2008time}:
	 \begin{equation}
		y_{c}<y_{b}+\left(y_{a}-y_{b}\right) \frac{t_{b}-t_{c}}{t_{b}-t_{a}}
	 \end{equation}

	\subsection{Superposed random walk}
	
	First, an undirected graph $G(V, E)$ is given. According to previous research \cite{zhang2017novel}, $G$ has a matrix $P$ recording transition probabilities and $P_{xy}$ is the probability that
	random walkers walking  from node $x$ to $y$ which define as  $P_{xy}=\frac{a_{xy}}{k_x}$. Matrix $A$ is the adjacency matrix of $G$ \cite{liu2010link}.
	
	The probability vector  $\vec{\pi}_{x}(t)$ which records the probability of a walker walks to each node of $G$ in $t$ step from node $x$. When $t=0$, $\vec{\pi}_{x}(0)$ is initialized to a vector in the shape of $N \times 1$ which $x$-th element is 1 and others equal to 0, $N=|V|$. And after $t$ steps, $\vec{\pi}_{x}(t)$ can calculate as follows:
	
	\begin{equation}
		\vec{\pi}_{x}(t)=P^{T}\vec{\pi}_{x}(t-1)
	\end{equation}

	Then the local similarity $S_{xy}^{LRW}$ between node $x$ and $y$ is calculated
	as follows:
	\begin{equation}
		 S^{LRW}_{xy}(t)=\frac{k_x}{2\left | E \right |}\times\vec{\pi}_{xy}(t) +\frac{k_y}{2\left | E \right |}\times\vec{\pi}_{yx}(t)
	 \end{equation}
	Superposed similarity $S^{SRW}$ is superposed local similarity $S^{LRW}$ until local similarities no longer change which defines as follows:
	\begin{equation}
		S^{SRW}_{xy}=\sum_{l=1}^{t}S^{LRW}_{xy}(l)
	\end{equation}

	\section{The proposed method}
	
		\subsection{Step 1: Transforming time series}
		Consider a time series $U = \left\{(t_1, y_1), (t_2, y_2),..., (t_n, y_n)\right\}$. The first step is to transform the time series $U$ into a undirected  graph $G(V, E)$ through the VG definition. 
		
		Following  Fig.1 at the is a histogram of a time series $U_1$ which defines as following Eq.6, and edges with visible node pairs have been marked with red segments.
		
		\begin{equation}
			\begin{aligned}
				U_1 = \left\{ (1,10),(2,90),(3,30),(4,50),(5,20), \right.\\
				\left.(6,40),(7,60),(8,50),(9,30),(10,40)\right\}
		\end{aligned}
	\end{equation}

	By calculating the visual relationship between each pair of nodes, the adjacency matrix $A_1$ which defined in Eq.7 of $G_1$ can be obtained. According to matrix $A_1$, the time series $U_1$ is transformed into $G_1$ as shown in Fig.2.
	\begin{equation}
			A_1 = \begin{bmatrix}
			0 & 1 & 0 & 0 & 0 & 0 & 0 & 0 & 0 & 0 \\
			1 & 0 & 1 & 1 & 0 & 1 & 1 & 0 & 0 & 0 \\
			0 & 1 & 0 & 1 & 0 & 0 & 0 & 0 & 0 & 0 \\
			0 & 1 & 1 & 0 & 1 & 1 & 1 & 0 & 0 & 0 \\
			0 & 0 & 0 & 1 & 0 & 1 & 1 & 0 & 0 & 0 \\
			0 & 1 & 0 & 1 & 1 & 0 & 1 & 0 & 0 & 0 \\
			0 & 1 & 0 & 1 & 1 & 1 & 0 & 1 & 0 & 1 \\
			0 & 0 & 0 & 0 & 0 & 0 & 1 & 0 & 1 & 1 \\
			0 & 0 & 0 & 0 & 0 & 0 & 0 & 1 & 0 & 1 \\
			0 & 0 & 0 & 0 & 0 & 0 & 1 & 1 & 1 & 0
		\end{bmatrix}
	\end{equation}

	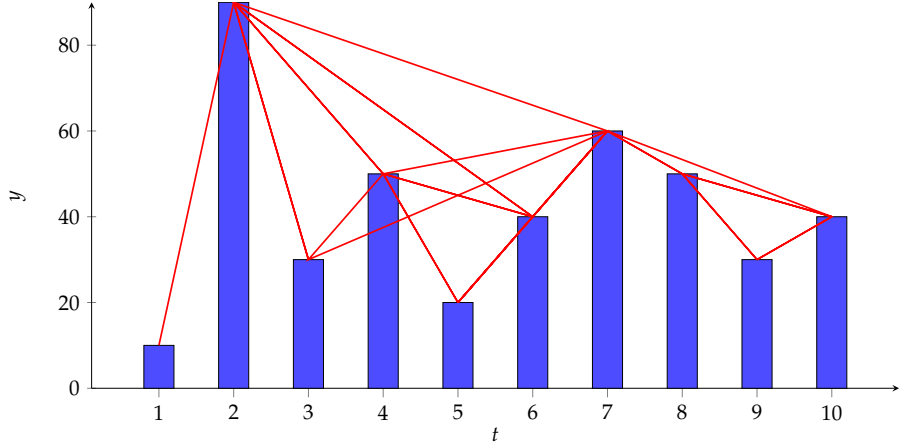
\begin{figure}[htbp]
		\centering
		\begin{tikzpicture}[scale = 0.8]
			\begin{axis}[
				axis lines=left,
				xlabel = $t$,
				ylabel = $y$,
				ybar=-0.5cm,
				bar width=0.5cm,
				ymin= 0,
				width=15cm,
				height=8cm,
				enlarge x limits=.1,
				]
				\addplot[draw=black,fill=blue!70] coordinates {(1, 10.0) (2, 90.0) (3, 30.0) (4, 50.0) (5, 20.0) (6, 40.0) (7, 60.0) (8, 50.0) (9, 30.0) (10, 40.0) };
				\addplot[sharp plot,thick, draw=red!100] coordinates {
					(1, 10.0) (2, 90.0) 
					(2, 90.0) (3, 30.0) (2, 90.0) (4, 50.0) (2, 90.0) (6, 40.0) (2, 90.0) (7, 60.0) 
					(3, 30.0) (4, 50.0) 
					(4, 50.0) (5, 20.0) (4, 50.0) (6, 40.0) (4, 50.0) (7, 60.0) 
					(5, 20.0) (6, 40.0) (5, 20.0) (7, 60.0) 
					(6, 40.0) (7, 60.0) 
					(7, 60.0) (8, 50.0) (7, 60.0) (10, 40.0) 
					(8, 50.0) (9, 30.0) (8, 50.0) (10, 40.0) 
					(9, 30.0) (10, 40.0) 
				};
			\end{axis}
		\end{tikzpicture}
			\caption{The primary time series $U_1$ of 10 time points}
\end{figure}
		
\begin{figure}[htbp]
	\centering
		\begin{tikzpicture}[scale = 1]
			\centering
			\tikzstyle{every node}=[draw,shape=circle]
			\draw[thick, draw=blue!50](1,0) node[circle,fill,inner sep=3pt,label=below: $t_{1}$ ](1){} -- (2,0) node[circle,fill,inner sep=3pt,label=below: $t_{2}$ ](2){};
			\draw[thick, draw=blue!50](2) -- (3,0) node[circle,fill,inner sep=3pt,label=below: $t_{3}$ ](3){};
			\draw[thick, draw=blue!50](3) -- (4,0) node[circle,fill,inner sep=3pt,label=below: $t_{4}$ ](4){};
			\draw[thick, draw=blue!50](4) -- (5,0) node[circle,fill,inner sep=3pt,label=below: $t_{5}$ ](5){};
			\draw[thick, draw=blue!50](5) -- (6,0) node[circle,fill,inner sep=3pt,label=below: $t_{6}$ ](6){};
			\draw[thick, draw=blue!50](6) -- (7,0) node[circle,fill,inner sep=3pt,label=below: $t_{7}$ ](7){};
			\draw[thick, draw=blue!50](7) -- (8,0) node[circle,fill,inner sep=3pt,label=below: $t_{8}$ ](8){};
			\draw[thick, draw=blue!50](8) -- (9,0) node[circle,fill,inner sep=3pt,label=below: $t_{9}$ ](9){};
			\draw[thick, draw=blue!50](9) -- (10,0) node[circle,fill,inner sep=3pt,label=below: $t_{10}$ ](10){};
			\draw[thick, draw=blue!50](2) to[out=70,in=120] (4,0);
			\draw[thick, draw=blue!50](2) to[out=70,in=120] (6,0);
			\draw[thick, draw=blue!50](2) to[out=70,in=120] (7,0);
			\draw[thick, draw=blue!50](4) to[out=70,in=120] (6,0);
			\draw[thick, draw=blue!50](4) to[out=70,in=120] (7,0);
			\draw[thick, draw=blue!50](5) to[out=70,in=120] (7,0);
			\draw[thick, draw=blue!50](7) to[out=70,in=120] (10,0);
			\draw[thick, draw=blue!50](8) to[out=70,in=120] (10,0);
		\end{tikzpicture}
		\caption{The transformed Visibility Graph $G_1$ of the time series}
	\end{figure}
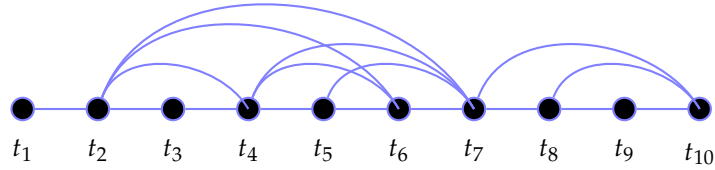

	\subsection{Step 2: Calculate node similarity}
	In $G$, there are many pairs of nodes, and each pair of nodes has similarity. According to previous studies \cite{zhang2017novel}, by finding the node $v'$ with the highest similarity to the last node $v$ of the time series $U$, the forecasting effect of linear fitting is positive.
	
	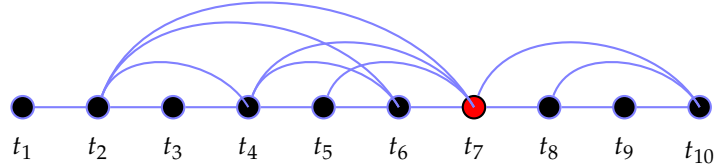
\begin{figure}[htbp]
		\centering
		\begin{tikzpicture}[scale = 1]
			\centering
			\tikzstyle{every node}=[draw,shape=circle]
			\draw[thick, draw=blue!50](1,0) node[circle,fill,inner sep=3pt,label=below: $t_{1}$ ](1){} -- (2,0) node[circle,fill,inner sep=3pt,label=below: $t_{2}$ ](2){};
			\draw[thick, draw=blue!50](2) -- (3,0) node[circle,fill,inner sep=3pt,label=below: $t_{3}$ ](3){};
			\draw[thick, draw=blue!50](3) -- (4,0) node[circle,fill,inner sep=3pt,label=below: $t_{4}$ ](4){};
			\draw[thick, draw=blue!50](4) -- (5,0) node[circle,fill,inner sep=3pt,label=below: $t_{5}$ ](5){};
			\draw[thick, draw=blue!50](5) -- (6,0) node[circle,fill,inner sep=3pt,label=below: $t_{6}$ ](6){};
			\draw[thick, draw=blue!50](6) -- (7,0) node[draw = black!100, circle,fill=red!100,inner sep=3pt,label=below: $t_{7}$ ](7){};
			\draw[thick, draw=blue!50](7) -- (8,0) node[circle,fill,inner sep=3pt,label=below: $t_{8}$ ](8){};
			\draw[thick, draw=blue!50](8) -- (9,0) node[circle,fill,inner sep=3pt,label=below: $t_{9}$ ](9){};
			\draw[thick, draw=blue!50](9) -- (10,0) node[circle,fill,inner sep=3pt,label=below: $t_{10}$ ](10){};
			\draw[thick, draw=blue!50](2) to[out=70,in=120] (4,0);
			\draw[thick, draw=blue!50](2) to[out=70,in=120] (6,0);
			\draw[thick, draw=blue!50](2) to[out=70,in=120] (7,0);
			\draw[thick, draw=blue!50](4) to[out=70,in=120] (6,0);
			\draw[thick, draw=blue!50](4) to[out=70,in=120] (7,0);
			\draw[thick, draw=blue!50](5) to[out=70,in=120] (7,0);
			\draw[thick, draw=blue!50](7) to[out=70,in=120] (10,0);
			\draw[thick, draw=blue!50](8) to[out=70,in=120] (10,0);
		\end{tikzpicture}
		\caption{The Zhang et al. 's method \cite{zhang2017novel} only uses the node $t_7$ which is the most similar to the last node $t_{10}$ for the final forecasting}
	\end{figure}

		\begin{figure}[htbp]
		\centering
		\begin{tikzpicture}[scale = 1]
			\centering
			\tikzstyle{every node}=[draw,shape=circle]
			\draw[thick, draw=blue!50](1,0) node[draw = black!100, circle,fill=red!10,inner sep=3pt,label=below: $t_{1}$ ](1){} -- (2,0) node[draw = black!100, circle,fill=red!80,inner sep=3pt,label=below: $t_{2}$ ](2){};
			\draw[thick, draw=blue!50](2) -- (3,0) node[draw = black!100, circle,fill=red!30,inner sep=3pt,label=below: $t_{3}$ ](3){};
			\draw[thick, draw=blue!50](3) -- (4,0) node[draw = black!100, circle,fill=red!70,inner sep=3pt,label=below: $t_{4}$ ](4){};
			\draw[thick, draw=blue!50](4) -- (5,0) node[draw = black!100, circle,fill=red!50,inner sep=3pt,label=below: $t_{5}$ ](5){};
			\draw[thick, draw=blue!50](5) -- (6,0) node[draw = black!100, circle,fill=red!60,inner sep=3pt,label=below: $t_{6}$ ](6){};
			\draw[thick, draw=blue!50](6) -- (7,0) node[draw = black!100, circle,fill=red!90,inner sep=3pt,label=below: $t_{7}$ ](7){};
			\draw[thick, draw=blue!50](7) -- (8,0) node[draw = black!100, circle,fill=red!40,inner sep=3pt,label=below: $t_{8}$ ](8){};
			\draw[thick, draw=blue!50](8) -- (9,0) node[draw = black!100, circle,fill=red!20,inner sep=3pt,label=below: $t_{9}$ ](9){};
			\draw[thick, draw=blue!50](9) -- (10,0) node[draw = black!100, circle,fill=blue!50,inner sep=3pt,label=below: $t_{10}$ ](10){};
			\draw[thick, draw=blue!50](2) to[out=70,in=120] (4,0);
			\draw[thick, draw=blue!50](2) to[out=70,in=120] (6,0);
			\draw[thick, draw=blue!50](2) to[out=70,in=120] (7,0);
			\draw[thick, draw=blue!50](4) to[out=70,in=120] (6,0);
			\draw[thick, draw=blue!50](4) to[out=70,in=120] (7,0);
			\draw[thick, draw=blue!50](5) to[out=70,in=120] (7,0);
			\draw[thick, draw=blue!50](7) to[out=70,in=120] (10,0);
			\draw[thick, draw=blue!50](8) to[out=70,in=120] (10,0);
		\end{tikzpicture}
		\caption{The proposed method retains all the similarity information $S$ of the last node $t_{10}$ (the color of nodes $t_1-t_9$ shows all the similarity of the last node $t_{10}$, and the more red the more similar).}
	\end{figure}
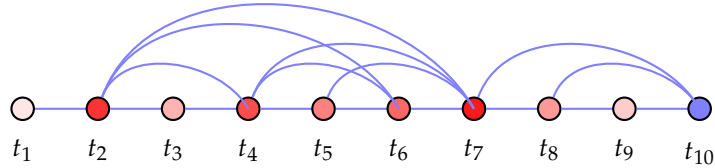
	
	This paper improves on previous studies \cite{zhang2017novel, mao2019time}. Fig.3 shows the use of node similarity in forecasting by Zhang et al. 's forecasting method \cite{zhang2017novel}. It can be seen that only one node $t_7$ is used. However, as the number of nodes in the time series becomes more, a weighting method is considered for forecasting. Through the SRW algorithm, the similarity between the previous node and the last node $v$ forms a similarity vector $S = [s_1, s_2, ...., s_{N-1}]$, $s_i = S_{iN}^{SRW}$ represents superposed similarity between node $i$ and node $n$ , $N$ is the number of nodes in the time series $U$. As shown in Fig.4, the proposed method retains all the similarity information and uses all the node information in the forecasting.
	
	According to previous studies \cite{zhang2017novel}, each node $i$ has a corresponding single step predicted value $\hat{y_i}$.  $\hat{y_i}$ is the predicted value by linear fitting node $i$ to node $v$, as shown in Fig.5. $\hat{y_i}$ is calculated in Eq.8. Also, $y_{v'}$ is the predicted value of the previous research \cite{zhang2017novel}. 
	
\begin{figure}[tbph]
	\centering
	\centerline{\includegraphics[scale=0.6]{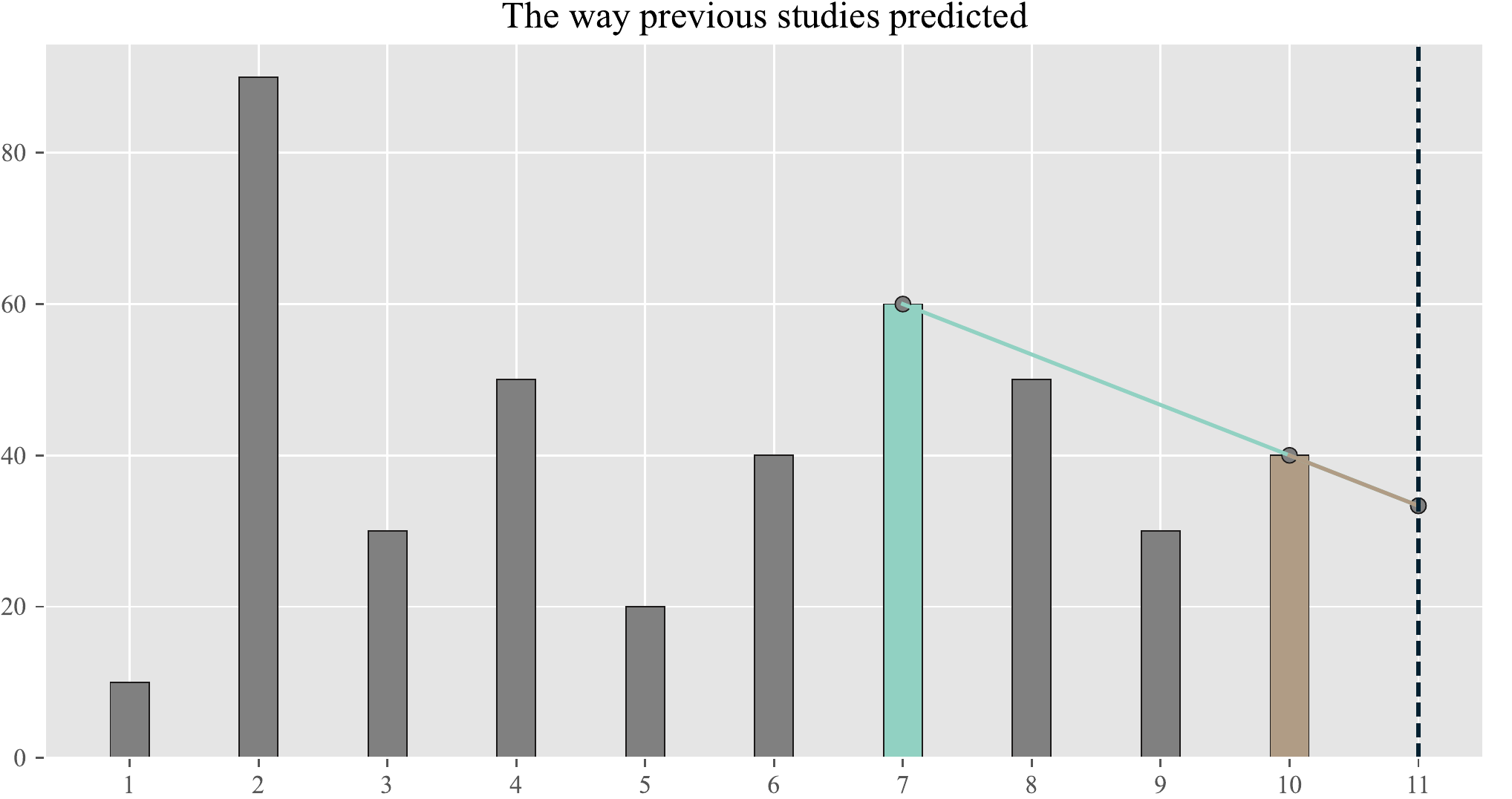}}
	\caption{Schematic diagram of Zhang et al. 's method \cite{zhang2017novel}, and node $t_{7}$ is the node most similar to the last node $t_{10}$}
\end{figure}

\begin{figure}[tbph]
	\centering
	\centerline{\includegraphics[scale=0.6]{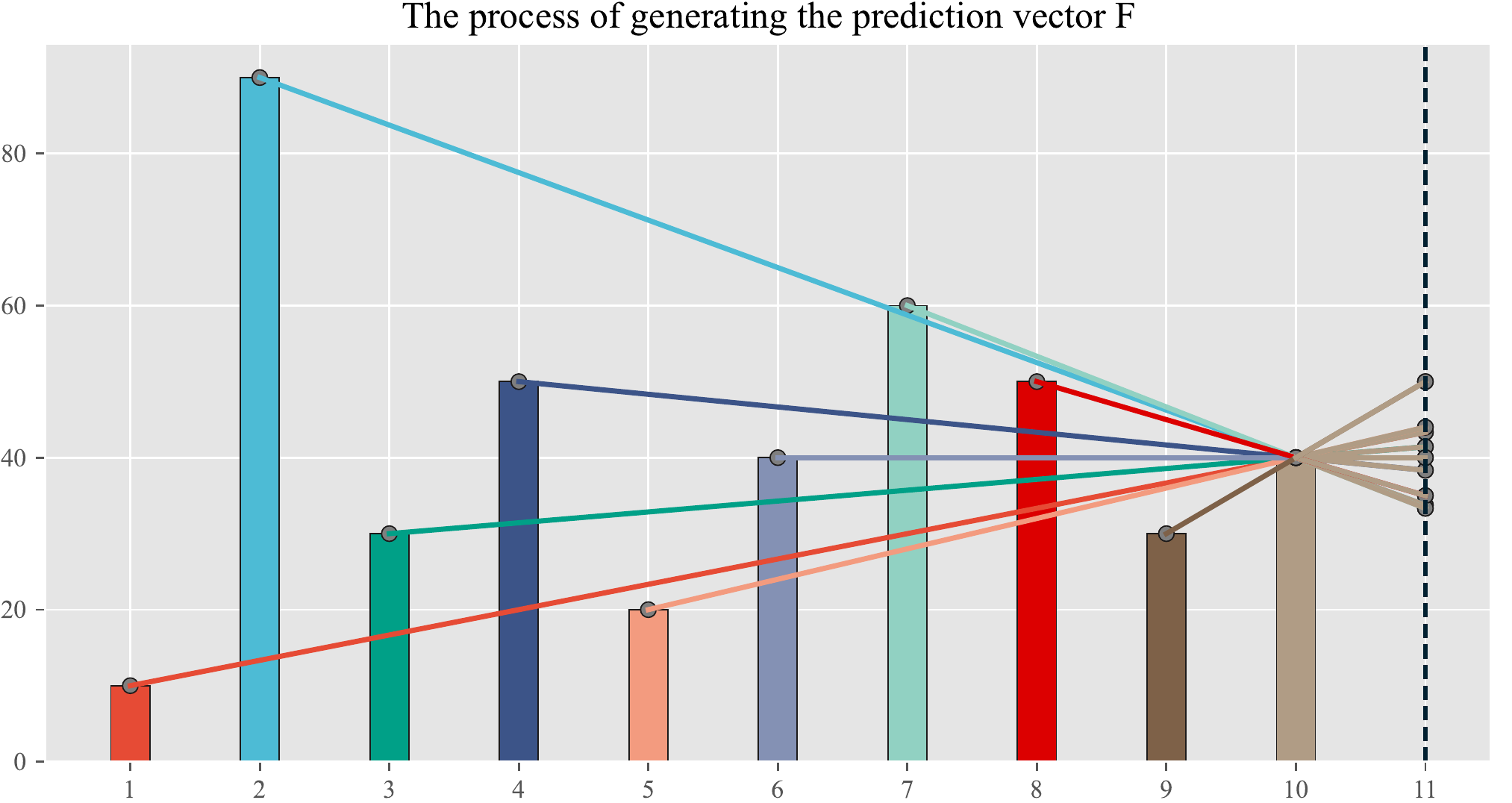}}
	\caption{The process that produces the forecasting vector $F_1$, the nodes' value on $x=11$ is the value of the vector $F_1$.}
\end{figure}
	
	\begin{equation}
	 \hat{y_i} = y_N + \frac{y_N-y_i}{t_N - t_i}
	\end{equation}

	However, when the number of nodes in the time series increases, with the influence of randomness and uncertainty, the single most similar node will lead to the decline of the forecasting effect. The proposed method first presents a forecasting vector $F$. Eq.10 is the forecasting vector $F_1$ of time series $U_1$ and Fig.6 shows the process of calculating vector $F_1$.
	
	\begin{mydef}
		Forecasting vector $F$ records the predicted value of all nodes except the last one as Eq.9.
		\begin{equation}
			F = \left\{\hat{y}_1, \hat{y}_2, ..., \hat{y}_{N-1} \right\}
		\end{equation}
	\end{mydef}

	\begin{equation}
		F_1 = [43.33, 33.75, 41.42, 38.33, 44.00, 40.00, 33.33, 35.00, 50.00]
	\end{equation}

	Given the validity of previous research, here the weight vector $w$ will come from the vector $S$. 
	
	\begin{mydef}
		In the proposed method, the weight vector $w$ is the normalized similarity vector $S$. For each element $s_i$ in $S$, it is standardized according to $w_i = s_i = \frac{s_i}{\sum_{l=1}^{N-1} s_l}$, so that the two-norm of S is 1, which meets the condition of being a weight.
	\end{mydef}

	\subsection{Step 3: Weighted forecasting vector}
		The previous study is equivalent to assigning all the weights to the nodes with the largest similarity, that is, all the weights in $w$ are $0$ except the position of $argmax (S)$ is 1. However, there is often a situation that the number of nodes with the largest similarity is not unique, and in this case, multiple predicted values will be generated. Therefore, the proposed method is to take the similarity as the weight vector $w$, and the weighted average of the forecasting vector $F$ as the forecasting value $\hat{y}$. The predicted value $\hat{y}$ is the dot product of the weight vector $w$ and the predicted vector $F$, as shown in Eq.11.
		
		\begin{equation}
			\hat{y} = w \cdot  F
		\end{equation}
	
		In the time series $U_1$, the weight vector of Zhang et al. 's \cite{zhang2017novel} method is as in Eq.12, and only position of node $t_7$ is 1. The proposed similarity vector $S_{Proposed}$ and weight vector $w_{Proposed}$ are shown in Eq.13, Eq.14. The difference with Zhang et al. 's method \cite{zhang2017novel} is that the weight of the proposed method retains the similarity information of nodes $t_1-t_9$.
		
		\begin{equation}
			w_{Zhang} = [0, 0, 0 , 0, 0, 0, 1, 0 ,0]
		\end{equation}
		
		\begin{equation}
			\begin{aligned}
				S_{Proposed} =	& [0.2213, 1.1194, 0.4245, 1.0290,  0.6029, \\
				&0.8201, 1.1990, 0.5684, 0.3737]
			\end{aligned}
		\end{equation}
		
		\begin{equation}
			\begin{aligned}
				w_{Proposed} = &[0.0348, 0.1761, 0.0668, 0.1618, 0.0948, \\
				&0.1290, 0.1886, 0.0894, 0.0588] \left( \parallel w_{Proposed} \parallel ^2 = 1\right)
			\end{aligned}
		\end{equation}
	
		Thus, the predicted values of the proposed method and Zhang et al. 's method \cite{zhang2017novel} are the same as shown in Eq.15, Eq.16. Of course, the time series $U_1$ is only an example, it is a simulated series, and the exact value at node $t_{11}$ cannot be determined. In the experiment section, the effectiveness of the proposed method will be compared. Algorithm 1 below is the pseudocode of the proposed method.
		
		\begin{equation}
				\hat{y}_{Zhang} = w_{Zhang}*F_1 = \hat{y}_7 =  33.33
		\end{equation}
		
		\begin{equation}
			\hat{y}_{Proposed} = w_{Proposed}*F_1 = \sum_{l=1}^{9} (w_l \times \hat{y}_l) = 38.10
		\end{equation}
	
			\begin{algorithm}[htbp]
			\caption{The proposed forecasting method}
			\begin{algorithmic}[1]
				\State Load time series $U$
				\For{Each node pairs $(i,j)$ in $U$}
				\State  Calculate $A_{ij}$
				\EndFor
				\State Calculate the transition matrix $P$
				\State Initialize  forecasting vector $F$
				\For{Each node $i$ in $U$ without last node}
				\State Calculate $F_i$
				\EndFor
				\State Initialize superposed similarity  vector $S_{SRW}$
				\While{True}
					\For{Each node $i$ in $U$}
						\If{$t=0$}
							\State Initialize $\pi_i(0)$
						\Else
							 \State $\pi_i(t) \leftarrow \pi_i(t+1)$
						 \EndIf
					\EndFor
					\State Calculate local similarity vector $S_{LRW}$
					\State $S_{SRW} = S_{SRW} + S_{LRW}$
					\State Backup current $S_{LRW}$ as $S_{LRW}'$
					\If{ $t \neq 1$ and $S_{LRW} = S_{LRW}'$}
						\State Break
					\EndIf
				\EndWhile
				\State Similarity vector $S = S_{SRW}$
				\State Normalize $S$ as weight vector $w$
				\State Forecasting Value $\hat{y} = w \cdot F$
				\Return  $\hat{y}$	
			\end{algorithmic}
		\end{algorithm}

	\section{Experiments}
	 	\subsection{Construction Cost Index}
			The Engineering Cost Record (ENR) publishes the Construction Cost Index (CCI) once a month \cite{ashuri2010time}. The CCI data of the construction industry is worthy of reference data, and many scholars in the construction industry have conducted research. A total of 295 CCI data values (CCI data sets from January 1990 to July 2014) are used for forecasting time series.
			
			Firstly, a fragment of length 11 of CCI is selected and named as time series $U_2$ as shown in Eq.17. 10 of these nodes are used for forecastings, and the last node in time is used as a target (underlined item in Eq.17). Visibility Graph $G$ of time series $U_2$ is shown in Fig. 7. 
			
			\begin{equation}
			\begin{aligned}
				U_2 = \left\{ (1, 5071), (2,5070), (3,5106), (4,5167), (5,5262), (6,5260),  \right. \\
					\left. (7,5252), (8,5230), (9,5255), (10, 5264), \underline{(11, 5278 )}    \right\} 
			\end{aligned}
			\end{equation}
		
			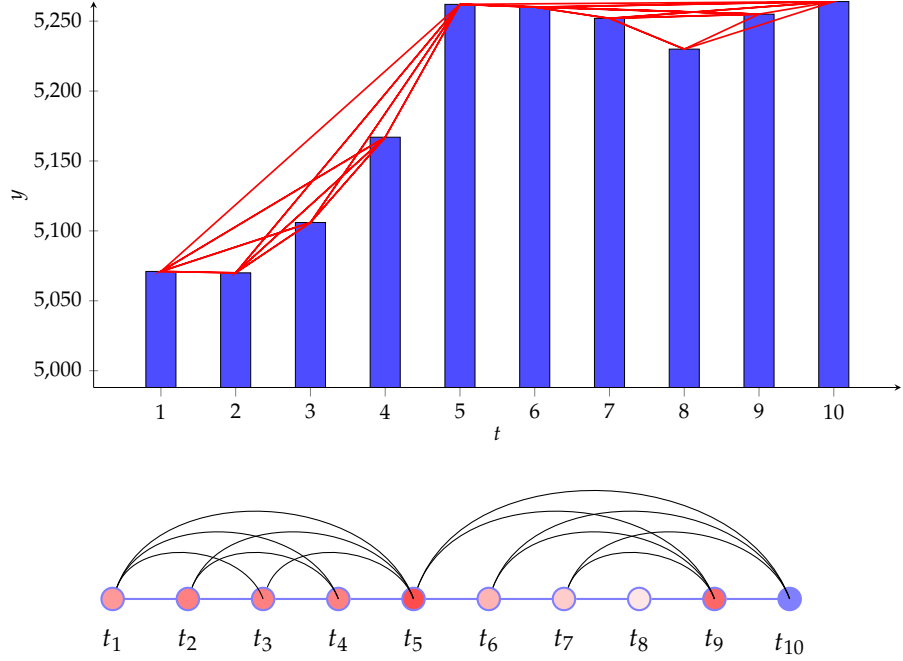
\begin{figure}[htbp]
				\centering
				
				\begin{tikzpicture}[scale = 0.8]
					\begin{axis}[
						axis lines=left,
						xlabel = $t$,
						ylabel = $y$,
						ybar=-0.5cm,
						bar width=0.5cm,
						ymin=4988,
						width=15cm, 
						height=8cm, 
						enlarge x limits=.1,
						]
						\addplot[draw=black,fill=blue!70] coordinates {(1, 5071.0) (2, 5070.0) (3, 5106.0) (4, 5167.0) (5, 5262.0) (6, 5260.0) (7, 5252.0) (8, 5230.0) (9, 5255.0) (10, 5264.0) };
						\addplot[sharp plot, thick,draw=red!100] coordinates {
							(1, 5071.0) (2, 5070.0) (1, 5071.0) (3, 5106.0) (1, 5071.0) (4, 5167.0) (1, 5071.0) (5, 5262.0) 
							(2, 5070.0) (3, 5106.0) (2, 5070.0) (4, 5167.0) (2, 5070.0) (5, 5262.0) 
							(3, 5106.0) (4, 5167.0) (3, 5106.0) (5, 5262.0) 
							(4, 5167.0) (5, 5262.0) 
							(5, 5262.0) (6, 5260.0) (5, 5262.0) (9, 5255.0) (5, 5262.0) (10, 5264.0) 
							(6, 5260.0) (7, 5252.0) (6, 5260.0) (9, 5255.0) (6, 5260.0) (10, 5264.0) 
							(7, 5252.0) (8, 5230.0) (7, 5252.0) (9, 5255.0) (7, 5252.0) (10, 5264.0) 
							(8, 5230.0) (9, 5255.0) 
							(9, 5255.0) (10, 5264.0) 
						};
					\end{axis}
				\end{tikzpicture}
				
				\begin{tikzpicture}[scale = 1]
					\centering
					\tikzstyle{every node}=[draw,shape=circle]
					\draw[thick, draw=blue!50] (1,0) node[circle,fill=red!40,inner sep=3pt,label=below: $t_{1}$ ](1){} -- (2,0) node[circle,fill=red!50,inner sep=3pt,label=below: $t_{2}$ ](2){};
					\draw[thick, draw=blue!50](2) -- (3,0) node[circle,fill=red!50,inner sep=3pt,label=below: $t_{3}$ ](3){};
					\draw[thick, draw=blue!50](3) -- (4,0) node[circle,fill=red!50,inner sep=3pt,label=below: $t_{4}$ ](4){};
					\draw[thick, draw=blue!50](4) -- (5,0) node[circle,fill=red!70,inner sep=3pt,label=below: $t_{5}$ ](5){};
					\draw[thick, draw=blue!50](5) -- (6,0) node[circle,fill=red!30,inner sep=3pt,label=below: $t_{6}$ ](6){};
					\draw[thick, draw=blue!50](6) -- (7,0) node[circle,fill=red!20,inner sep=3pt,label=below: $t_{7}$ ](7){};
					\draw[thick, draw=blue!50](7) -- (8,0) node[circle,fill=red!10,inner sep=3pt,label=below: $t_{8}$ ](8){};
					\draw[thick, draw=blue!50](8) -- (9,0) node[circle,fill=red!60,inner sep=3pt,label=below: $t_{9}$ ](9){};
					\draw[thick, draw=blue!50](9) -- (10,0) node[circle,fill = blue!50,inner sep=3pt,label=below: $t_{10}$ ](10){};
					\draw (1) to[out=70,in=110] (3,0);
					\draw (1) to[out=70,in=110] (4,0);
					\draw (1) to[out=70,in=110] (5,0);
					\draw (2) to[out=70,in=110] (4,0);
					\draw (2) to[out=70,in=110] (5,0);
					\draw (3) to[out=70,in=110] (5,0);
					\draw (5) to[out=70,in=110] (9,0);
					\draw (5) to[out=70,in=110] (10,0);
					\draw (6) to[out=70,in=110] (9,0);
					\draw (6) to[out=70,in=110] (10,0);
					\draw (7) to[out=70,in=110] (9,0);
					\draw (7) to[out=70,in=110] (10,0);
				\end{tikzpicture}
				\caption{ Histograms and visibility graph $G$ of time series $U_2$ (the color of nodes $t_1-t_9$ shows all the similarity of the last node $t_{10}$, and the more red the more similar).}
			\end{figure}
		
		The similarity vector $S_2$ corresponding to $Fig.7$ is shown in Eq.18. The similarity vector $S_2$ becomes weight vector $w_2$ after normalization, as shown in Eq.19. Fig.8 shows the generation process of the forecasting vector $F_2$.

		\begin{equation}
			S_2 = [0.8290, 0.8480, 0.8480, 0.8480, 1.4142, 0.7522, 0.7373, 0.3667, 0.9345]
		\end{equation}

			\begin{equation}
		\begin{aligned}
						w_2 = &[0.1094, 0.1119, 0.1119, 0.1119, 0.1866,\\
									& 0.0993, 0.0973, 0.0484, 0.1233] \left( \parallel w_{2} \parallel ^2 = 1\right)
		\end{aligned}
	\end{equation}
	
	\begin{figure}[tbph]
		\centering
		\centerline{\includegraphics[scale=0.6]{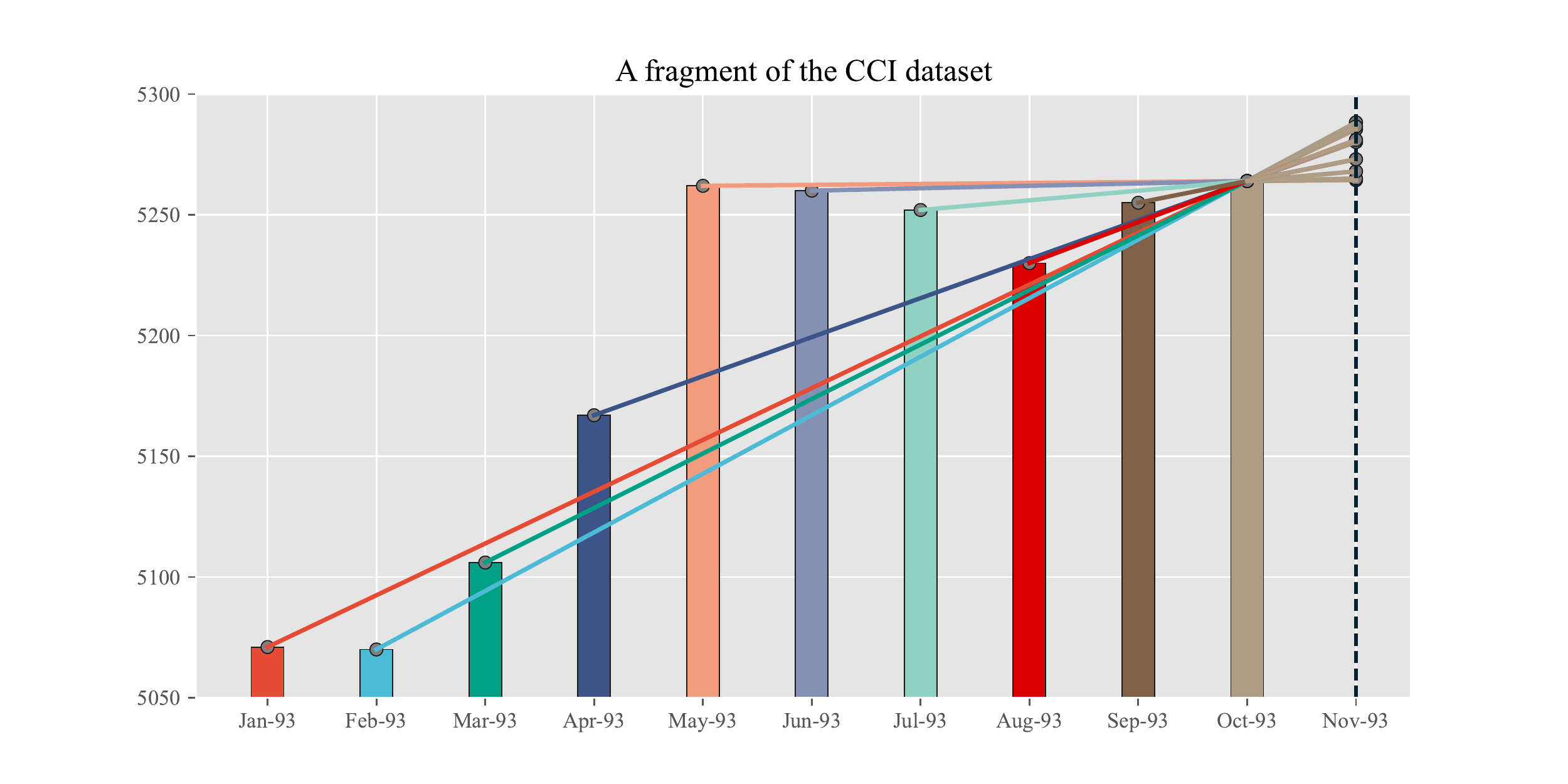}}
		\caption{The process that produces the forecasting vector $F_2$}
	\end{figure}
	
	\begin{equation}
		F_2 = [5285.44,5288.25,5286.57,5280.17, 5264.40,5265.00,5268.00, 5281.00, 5273.00]
	\end{equation}

	Through the forecasting vector $F_2$ and weight vector $w_2$, the predicted value of the proposed method $\hat{y}_{Proposed_2}$ and the predicted value of the previous work $\hat{y}_{Zhang_2}$ can be calculated, as shown in Tab.1. It can be seen that at node $t_{11}$, the proposed method is closer to the prediction target, and the error of previous research is larger. Of course, it is necessary to predict the whole time series and evaluate the prediction effect of the proposed method by using other comparison methods and error functions.
			
	\begin{table}[htbp]
		\centering
		\caption{Forecasting results of previous research and proposed methods on time series $U_2$ (numbers in parentheses indicate difference from target)}
		\begin{tabular}{ccc}
			\hline
			Target  & Zhang et al. \cite{zhang2017novel}    & Proposed      \\ \hline
			5278.00 & 5260.40(-17.6) & 5275.90(-2.1) \\ \hline
		\end{tabular}
	\end{table}
	
	In order to better complete the prediction, the CCI data were analyzed. Fig.9 is the original data of CCI, and the specific nature of CCI cannot be analyzed. The stationarity analysis of CCI dataset is carried out. Here, \textit{plot\_acf} and \textit{plot\_pacf} in Python package \textit{statsmodels.graphics.tsaplots} \cite{seabold2010statsmodels}  are used to draw autocorrelation and partial autocorrelation plots of CCI as shown in Fig.10. From Fig.10, the autocorrelation coefficient of CCI decreases slowly, so CCI is a non-stationary series. In addition, it can be seen from the partial autocorrelation coefficient that CCI data is only related to the former item.

	\begin{figure}[tbph]
		\centering
		\centerline{\includegraphics[scale=0.6]{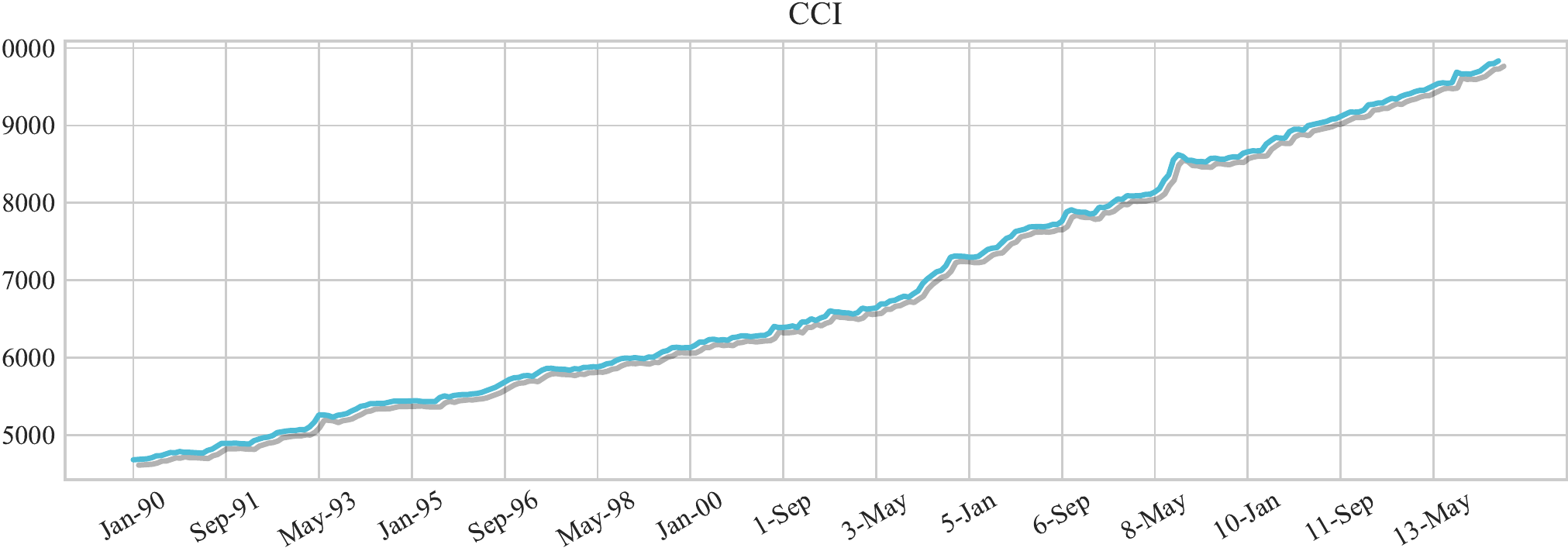}}
		\caption{CCI raw data}
	\end{figure}

		\begin{figure}[tbph]
		\centering
		\centerline{\includegraphics[scale=0.9]{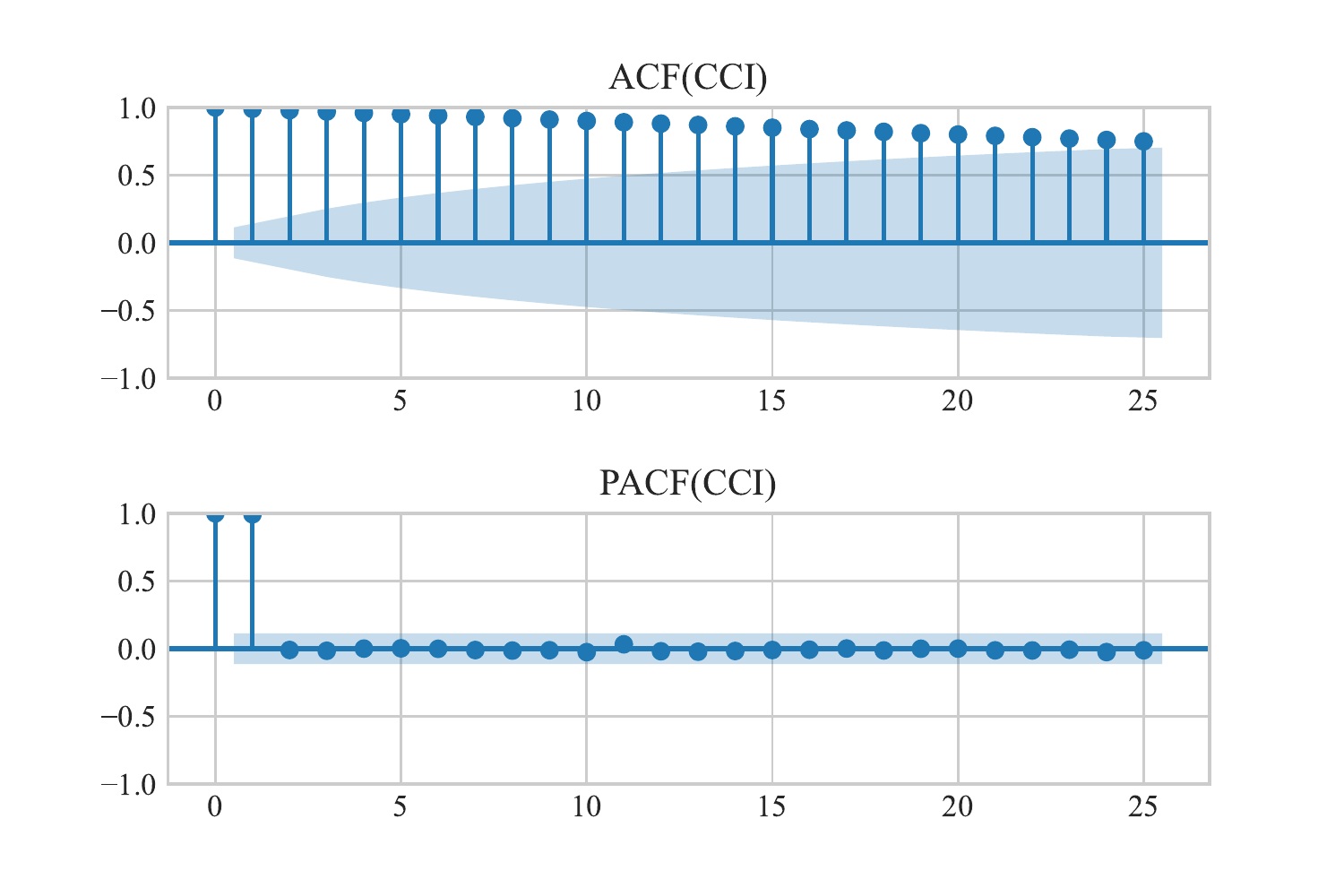}}
		\caption{Autocorrelation and partial autocorrelation plots of CCI}
	\end{figure}

	Therefore, the stationarity of CCI data is determined again by difference. Therefore, the stationarity of CCI data is determined again by difference. The autocorrelation coefficient of the first difference soon decays to near 0, which is obviously a stationary series.

	\begin{figure}[tbph]
		\centering
		\centerline{\includegraphics[scale=0.6]{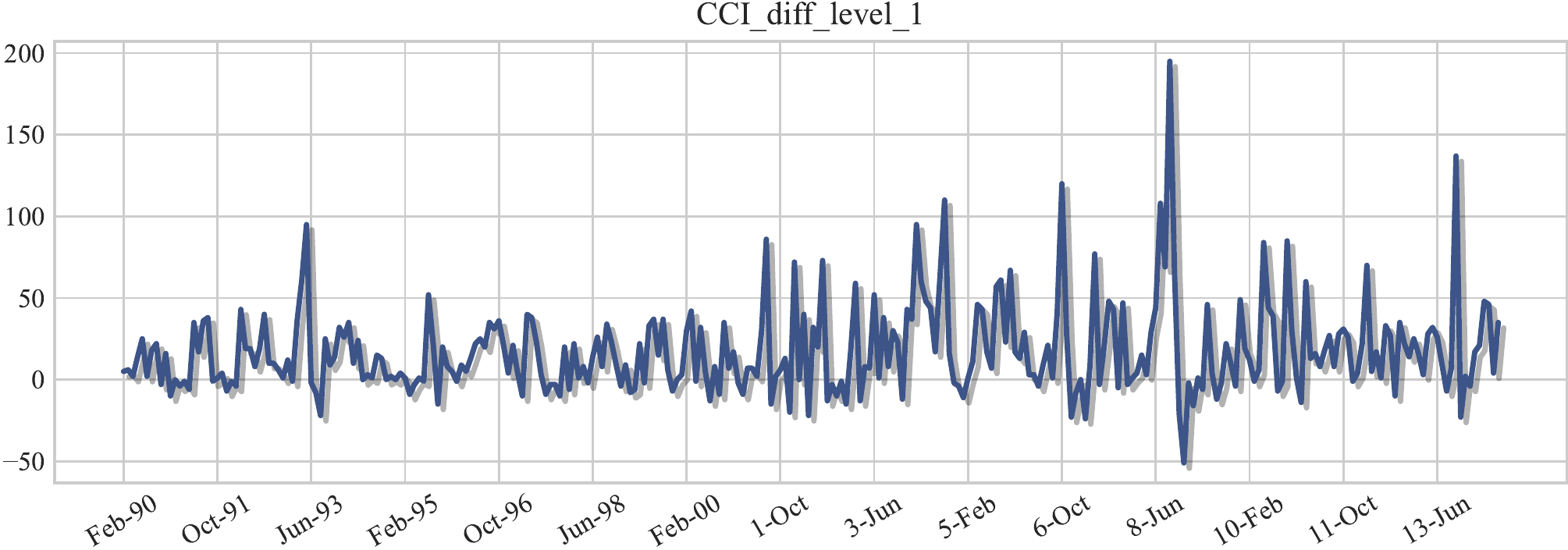}}
		\caption{ CCI of the first difference}
	\end{figure}

		\begin{figure}[tbph]
		\centering
		\centerline{\includegraphics[scale=0.9]{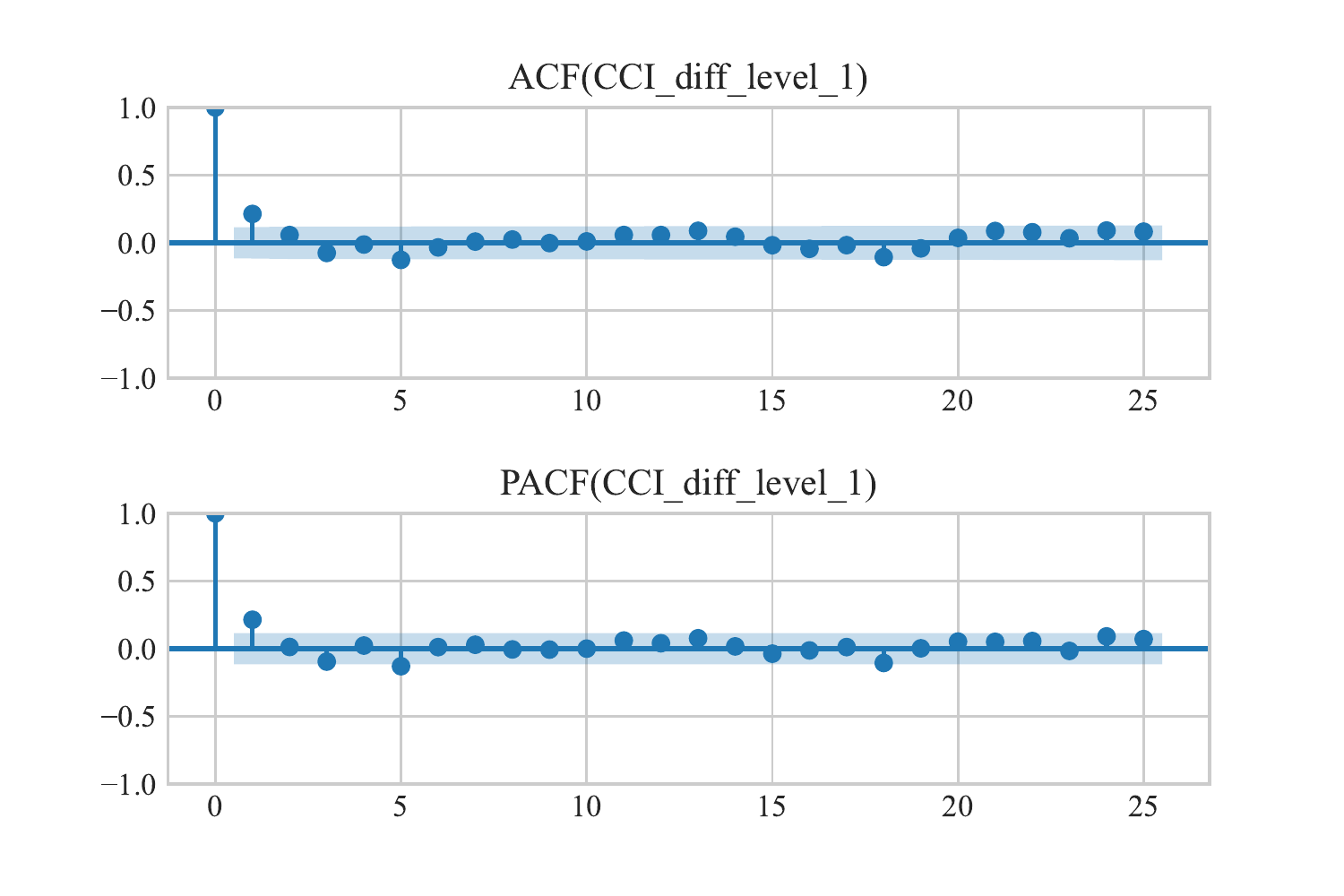}}
		\caption{Autocorrelation and partial autocorrelation of the first difference CCI}
	\end{figure}

	At the same time, the stationarity hypothesis test method is used to verify again. The current mainstream of hypothesis testing methods for stationarity is the unit root test, which tests whether there is a unit root in the series. If there is a unit root, it is a non-stationary series, and if there is no unit root, it is a stationary series. Dickey-fuller Test with GLS Detredding (DFGLS) is a unit root Test method proposed by Elliott, Rothenberg, and Stock \cite{baum2001dfgls}. The implementation of DFGLS is used here to verify CCI stationarity using \textit{DFGLS} from the Python \textit{ arch.unitroot} package \cite{kevin_sheppard_2022_6684078}.

	The test result is shown in Fig.13, the original data p-value$>0.05$, so the null hypothesis is not rejected, and CCI data is not stationary. According to CCI of the first difference's p-value$<0.05$, so the null hypothesis is rejected and the CCI data is stationary.  
	
	\begin{figure}[tbph]
		\centering
		\centerline{\includegraphics[scale=0.6]{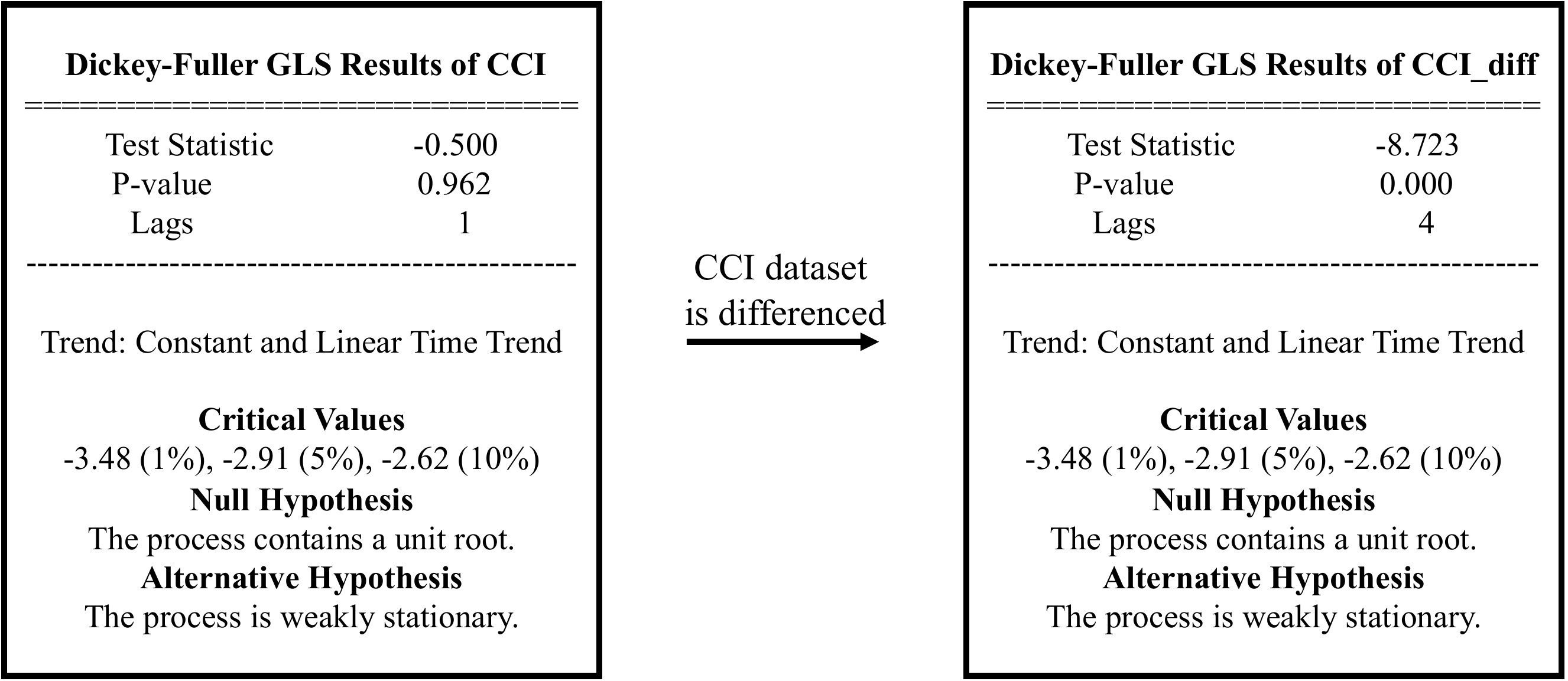}}
		\caption{Hypothesis test result}
	\end{figure}
	
	After data analysis, CCI will be predicted. Here, the first six data are taken as the retained data, and the prediction starts from the seventh data of CCI. Fig.14 is the prediction effect diagram of the proposed method and the comparison methods. The prediction results of the proposed method and the comparison method are both ideal. ARIMA in the comparison method is an implementation of \textit{auto\_arima} in the Python package \textit{pmdarmia.arima} \cite{nokeri2021forecasting, pmdarima63:online}. In order to more accurately compare the difference between the prediction results, mean absolute error (MAE), mean absolute percentage error (MAPE), Symmetric Mean Absolute Percentage Error (SMAPE), root mean square error (RMSE), and normalized root mean squared error (NRMSE) error indexes \cite{mao2019time} are used to measure, as shown in Eq.21 - Eq.25.
	
	\begin{equation}
	MAE=\frac{1}{N}\sum_{t=1}^{N}\left|\hat{y}(t)-y(t)\right|
	\end{equation}

	\begin{equation}
	MAPE=\frac{1}{N}\sum_{t=1}^{N}\frac{\left|\hat{y}(t)-y(t)\right|}{y(t)} 
\end{equation}

	\begin{equation}
	SMAPE=\frac{2}{N}\sum_{t=1}^{N}\frac{\left|\hat{y}(t)-y(t)\right|}{\hat{y}(t)+y(t)}
\end{equation}

	\begin{equation}
	RMSE=\sqrt{\frac{1}{N}\sum_{t=1}^{N}\left|\hat{y}(t)-y(t)\right|^{2}} 
\end{equation}

	\begin{equation}
	NRMSE=\frac{\sqrt{\frac{1}{N}\sum_{t=1}^{N}\left|\hat{y}(t)-y(t)\right|^{2}}}{y_{max}-y_{min}}
\end{equation}

	\begin{figure}[tbph]
		\centerline{\includegraphics[scale=0.6]{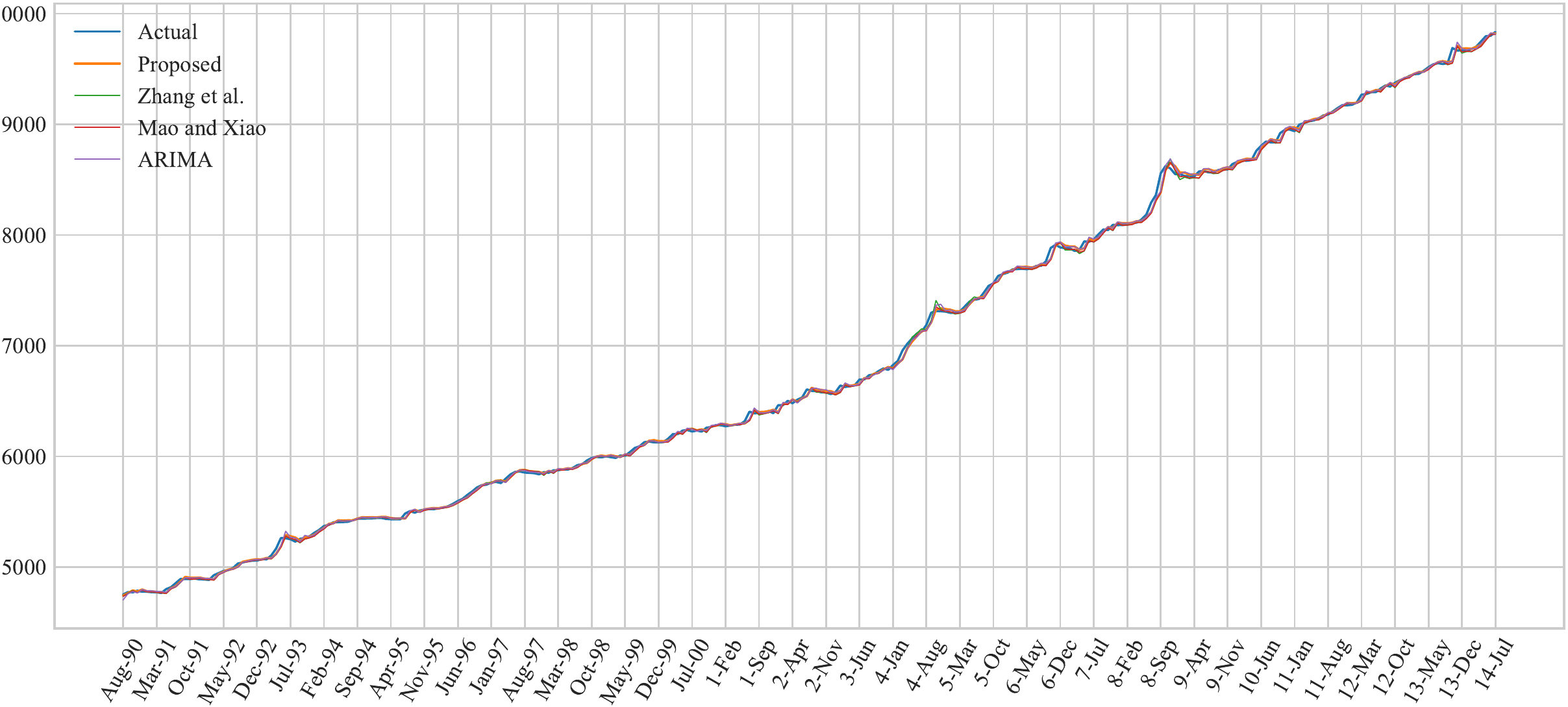}}
		\caption{The result of the proposed method and the comparison methods on CCI}
	\end{figure}

	Tab.2 shows the prediction errors of the proposed method and the comparison methods. According to Tab.2, the prediction error of the proposed method is the smallest, which is a great progress compared with previous studies in MAE index. At the same time, the statistical method ARIMA has an ideal prediction effect, but the error is larger than the proposed method. The proposed method will continue to validate the prediction effect on several competition datasets.
	
	\begin{table}[htbp]
		\centering
		\caption{The error of the proposed method and the comparison methods on CCI}
		\begin{tabular}{cccccc}
			\hline
			& \textbf{MAE}         & \textbf{MAPE}        & \textbf{SMAPE}       & \textbf{RMSE}        & \textbf{NRMSE}       \\ \hline
			Zhang et al. \cite{zhang2017novel}      & 21.10        & 0.3025          & 0.3031          & 30.74          & 43.12       \\
			Mao and Xiao \cite{mao2019time}     & 20.42         & 0.2936         & 0.2943           & 30.02          & 42.10         \\
			ARIMA \cite{ pmdarima63:online}           & 20.50       & 0.2971         & 0.2974          & 28.80          & 40.40          \\
			\textbf{Proposed} & \textbf{19.67} & \textbf{0.2847} & \textbf{0.2850} & \textbf{27.75} & \textbf{38.92} \\ \hline
		\end{tabular}
	\end{table}

		\begin{figure}[tbph]
		\centerline{\includegraphics[scale=0.35]{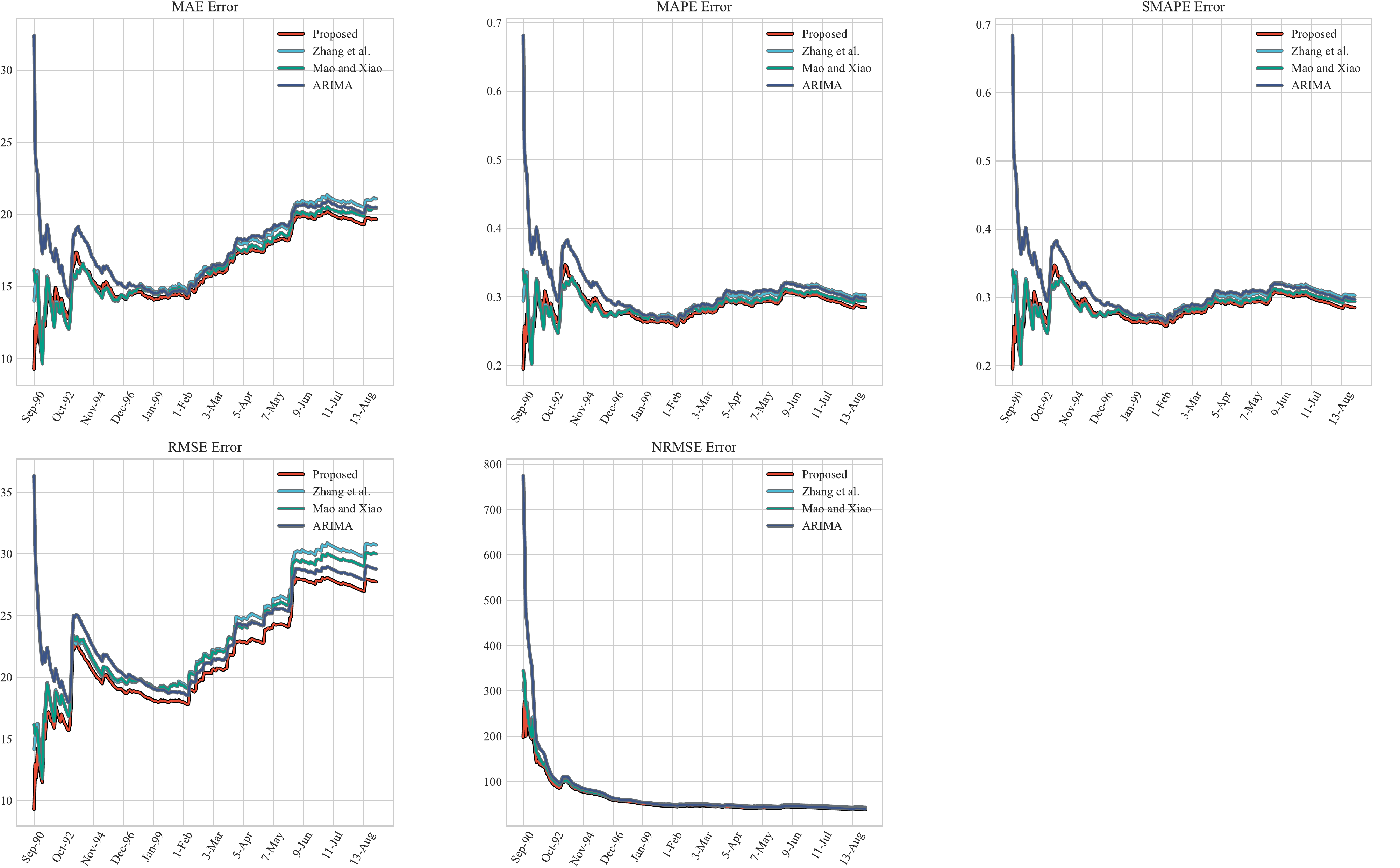}}
		\caption{Robustness verification: A record of each prediction}
	\end{figure}

		Meanwhile, in order to verify the robustness of the proposed method, the prediction errors of each time series are calculated. Fig.15 is the comparison of the prediction errors of the proposed method and the comparison methods at each time nodes. The error curve in red is the error curve of the proposed method. It can be seen that the error curve of the proposed method is below the curve of the comparison methods at most time points, indicating that the proposed method has small prediction error and strong robustness, and can maintain an ideal prediction effect in the continuous prediction.

	 	\subsection{Competition datasets}
	 	
	 	\begin{figure}[tbph]
	 		\centerline{\includegraphics[scale=0.25]{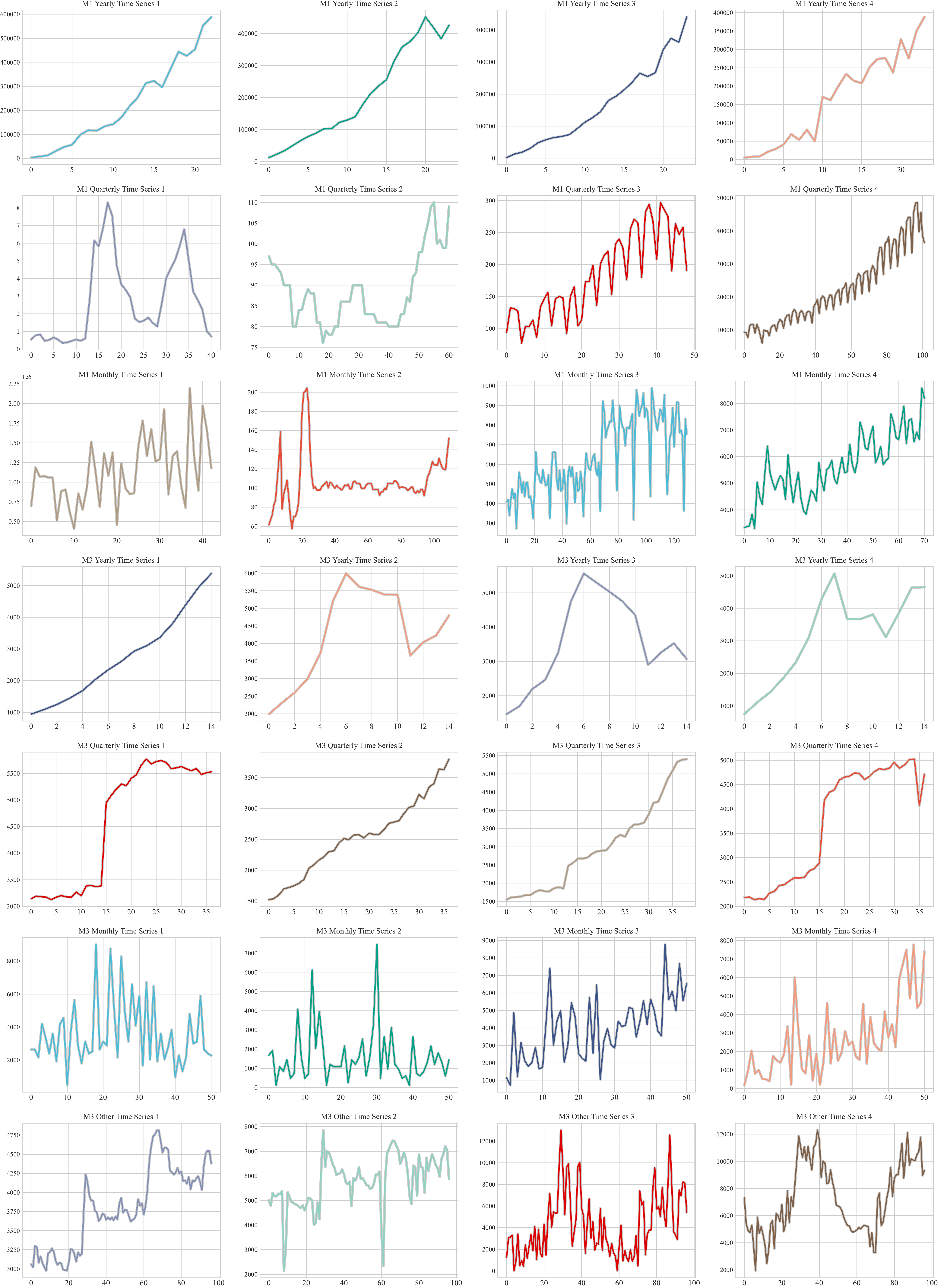}}
	 		\caption{Partial time series in M1,M3 data sets}
	 	\end{figure}
	 	
	 	In order to further verify the superiority of the proposed method, the proposed method will be validated in the competition dataset \cite{godahewa2021monash}. The Makridakis Competition (also known as the M Competition) is a series of open competitions to evaluate and compare the accuracy of different time series forecasting methods \cite{makridakis1979accuracy}. They were organized by a team led by forecasting researcher Spyros Makridakis and were first held in 1982. M1 \cite{makridakis1979accuracy, godahewa_rakshitha_2020_4656193, godahewa_rakshitha_2020_4656154, godahewa_rakshitha_2020_4656159} and M3 \cite{makridakis2000m3, godahewa_rakshitha_2020_4656222, godahewa_rakshitha_2020_4656262, godahewa_rakshitha_2020_4656298, godahewa_rakshitha_2020_4656335} data sets are used this time which shown in Fig.16, and the descriptive characteristics of the data sets \cite{godahewa2021monash} are shown in Tab.3.
	 	
	 	\begin{table}[htbp]
	 		\centering
	 		\caption{Feature of the competition dataset M1, M3}
	 		\begin{tabular}{cccc}
	 			\hline
	 			Dataset & Number of Series & Minimum Length & Maximum Length \\ \hline
	 			M1      & 1001             & 15             & 150            \\
	 			M3      & 3003             & 20             & 144            \\ \hline
	 		\end{tabular}
	 	\end{table}
	 	
	 	To illustrate the proposed method accuracy, the proposed method compared with methods including Simple Exponential Smoothing (SES) \cite{ostertagova2011simple}, Theta \cite{assimakopoulos2000theta}, Trigonometric Box-Cox ARMA Trend Seasonal Model (TBATS) \cite{de2011forecasting}, Exponential Smoothing (ETS) \cite{hyndman2008forecasting}, ARIMA \cite{box2015time}, Pooled Regression Model (PR) \cite{trapero2015identification}, CatBoost \cite{prokhorenkova2018catboost}, Feed-Forward Neural Network (FFNN) \cite{goodfellow2016deep}, DeepAR \cite{salinas2020deepar}, N-BEATS \cite{oreshkin2019n}, WaveNet \cite{borovykh2017conditional}, Transformer \cite{vaswani2017attention}. The proposed method is compared with the comparison methods in MAE and SMAPE. There are more than one time series in M1 and M3 data sets, so the average error of data sets is compared in the experiment. The experimental error in MAE is shown in Tab.4, and the experimental error in SMAPE is shown in Tab.5.

\begin{table}[htbp]
	\centering
	\setlength{\tabcolsep}{0.4mm}
	\caption{Mean MAE error of proposed method and comparison methods}
	\resizebox{\textwidth}{!}{
	\begin{tabular}{cccccccc}
		\hline
		Dataset      & SES \cite{ostertagova2011simple}      & Theta \cite{assimakopoulos2000theta}    & TBATS \cite{de2011forecasting}   & ETS  \cite{hyndman2008forecasting}      & ARIMA \cite{box2015time} & PR \cite{trapero2015identification} & CatBoost \cite{prokhorenkova2018catboost} \\ \hline
		M1-Yearly \cite{godahewa_rakshitha_2020_4656193}    & 171353.41 & 152799.26 & 103006.90 & 146110.11 & 145608.87   & 134246.38         & 215904.20 \\
		M1-Quarterly \cite{godahewa_rakshitha_2020_4656154} & 2206.27   & 1981.96   & 2326.46  & 2088.15   & 2191.10      & 1630.38           & 1802.18  \\
		M1-Monthly \cite{godahewa_rakshitha_2020_4656159}  & 2259.04   & 2166.18   & 2237.50   & 1905.28   & 2080.13     & 2088.25           & 2052.32  \\
		M3-Yearly \cite{godahewa_rakshitha_2020_4656222}   & 1022.27   & 957.40     & 1192.85  & 1031.40    & 1416.31     & 1018.48           & 1163.36  \\
		M3-Quarterly \cite{godahewa_rakshitha_2020_4656262} & 571.96    & 486.31    & 561.77   & 513.06    & 559.40       & 519.30             & 593.29   \\
		M3-Monthly \cite{godahewa_rakshitha_2020_4656298}  & 743.41    & 623.71    & 630.59   & 626.46    & 654.80       & 692.97            & 732.00      \\
		M3-Other \cite{godahewa_rakshitha_2020_4656335}     & 277.83    & 215.35    & 189.42   & 194.98    & 193.02      & 234.43            & 318.13   \\ \hline
		Dataset      & FFNN \cite{goodfellow2016deep}     & DeepAR \cite{salinas2020deepar}   & N-BEATS \cite{oreshkin2019n}  & WaveNet \cite{borovykh2017conditional}  & Transformer \cite{vaswani2017attention} & Proposed          &          \\ \cline{1-7}
		M1-Yearly \cite{godahewa_rakshitha_2020_4656193}   & 136238.80  & 152084.40  & 173300.20 & 284953.90  & 164637.90    & \textbf{68196.39} &          \\
		M1-Quarterly \cite{godahewa_rakshitha_2020_4656154} & 1617.39   & 1951.14   & 1820.25  & 1855.89   & 1864.08     & \textbf{1393.17}  &          \\
		M1-Monthly \cite{godahewa_rakshitha_2020_4656159}  & 2162.58   & 1860.81   & \textbf{1820.37}  & 2184.42   & 2723.88     & 2007.33  &          \\
		M3-Yearly  \cite{godahewa_rakshitha_2020_4656222}  & 1082.03   & 994.72    & 962.33   & 987.28    & 924.47      & \textbf{514.40}    &          \\
		M3-Quarterly \cite{godahewa_rakshitha_2020_4656262} & 528.47    & 519.35    & 494.85   & 523.04    & 719.62      & \textbf{406.19}   &          \\
		M3-Monthly \cite{godahewa_rakshitha_2020_4656298}  & 692.48    & 728.81    & 648.60    & 699.30     & 798.38      & \textbf{580.32}   &          \\
		M3-Other \cite{godahewa_rakshitha_2020_4656335}     & 240.17    & 247.56    & 221.85   & 245.29    & 239.24      & \textbf{91.80}     &          \\ \cline{1-7}
	\end{tabular}}
\end{table}

\begin{table}[htbp]
	\centering
	\setlength{\tabcolsep}{0.4mm}
	\caption{Mean SMAPE error of proposed method and comparison methods}
	\resizebox{\textwidth}{!}{
	\begin{tabular}{cccccccc}
		\hline
		Dataset      & SES \cite{ostertagova2011simple}  & Theta \cite{assimakopoulos2000theta}  & TBATS \cite{de2011forecasting}  & ETS \cite{hyndman2008forecasting}    & ARIMA \cite{box2015time} & PR \cite{trapero2015identification}               & CatBoost \cite{prokhorenkova2018catboost} \\ \hline
		M1-Yearly \cite{godahewa_rakshitha_2020_4656193}    & 23.10  & 20.17  & 17.42   & 18.61   & 19.47       & 18.79             & 20.25    \\
		M1-Quarterly \cite{godahewa_rakshitha_2020_4656154} & 18.10  & 16.35  & 16.65   & 17.47   & 16.62       & 16.67             & 17.60     \\
		M1-Monthly \cite{godahewa_rakshitha_2020_4656159}   & 17.43 & 16.53  & 15.15   & 15.05   & 15.65       & 15.20              & 16.51    \\
		M3-Yearly \cite{godahewa_rakshitha_2020_4656222}    & 17.76 & 16.76  & 17.37   & 17.00      & 18.84       & 17.13             & 20.07    \\
		M3-Quarterly \cite{godahewa_rakshitha_2020_4656262} & 10.90  & 9.20    & 10.22   & 9.68    & 10.24       & 9.77              & 11.18    \\
		M3-Monthly \cite{godahewa_rakshitha_2020_4656298}   & 16.22 & 13.86  & 13.85   & 14.14   & 14.24       & 15.17             & 16.41    \\
		M3-Other \cite{godahewa_rakshitha_2020_4656335}    & 6.28  & 4.92   & 4.35    & 4.37    & 4.35        & 5.32              & 6.74     \\ \hline
		Dataset      & FFNN \cite{goodfellow2016deep}     & DeepAR \cite{salinas2020deepar}   & N-BEATS \cite{oreshkin2019n}  & WaveNet \cite{borovykh2017conditional}  & Transformer \cite{vaswani2017attention} & Proposed          &          \\ \cline{1-7}
		M1-Yearly \cite{godahewa_rakshitha_2020_4656193}    & 18.20  & 18.72  & 20.52   & 21.25   & 18.96       & \textbf{8.73}     &          \\
		M1-Quarterly \cite{godahewa_rakshitha_2020_4656154} & 16.45 & 16.17  & 16.76   & 15.76   & 19.37       & \textbf{11.51}    &          \\
		M1-Monthly \cite{godahewa_rakshitha_2020_4656159}  & 15.71 & 17.16  & 16.77   & 16.59   & 22.17       & \textbf{14.65}    &          \\
		M3-Yearly \cite{godahewa_rakshitha_2020_4656222}   & 17.59 & 17.24  & 17.03   & 16.98   & 15.82       & \textbf{9.63}     &          \\
		M3-Quarterly \cite{godahewa_rakshitha_2020_4656262} & 9.90   & 9.93   & 9.47    & 9.72    & 13.17       & \textbf{8.10}      &          \\
		M3-Monthly \cite{godahewa_rakshitha_2020_4656298}  & 15.33 & 15.74  & 14.76   & 15.43   & 17.13       & \textbf{13.98}    &          \\
		M3-Other  \cite{godahewa_rakshitha_2020_4656335}   & 5.41  & 5.57   & 4.97    & 5.09    & 5.78        & \textbf{2.21}     &          \\ \cline{1-7}
	\end{tabular}}
\end{table}

	According to Tab.4 and Tab.5, the prediction accuracy of the proposed method is higher than that of the comparison method. Among the MAE errors, the proposed method is M3-yearly 514.40, and the comparison methods are all higher than 900, and the error of the proposed method is only about half of the comparison methods. However, the proposed method does not achieve the optimal performance in M1-monthly, because the short length of time series in M1-monthly does not reflect the advantages of the proposed method. Among the MAE errors, the accuracy of the proposed method exceeds that of all comparison methods on all data sets.
	
	\begin{figure}[tbph]
		\centerline{\includegraphics[scale=0.5]{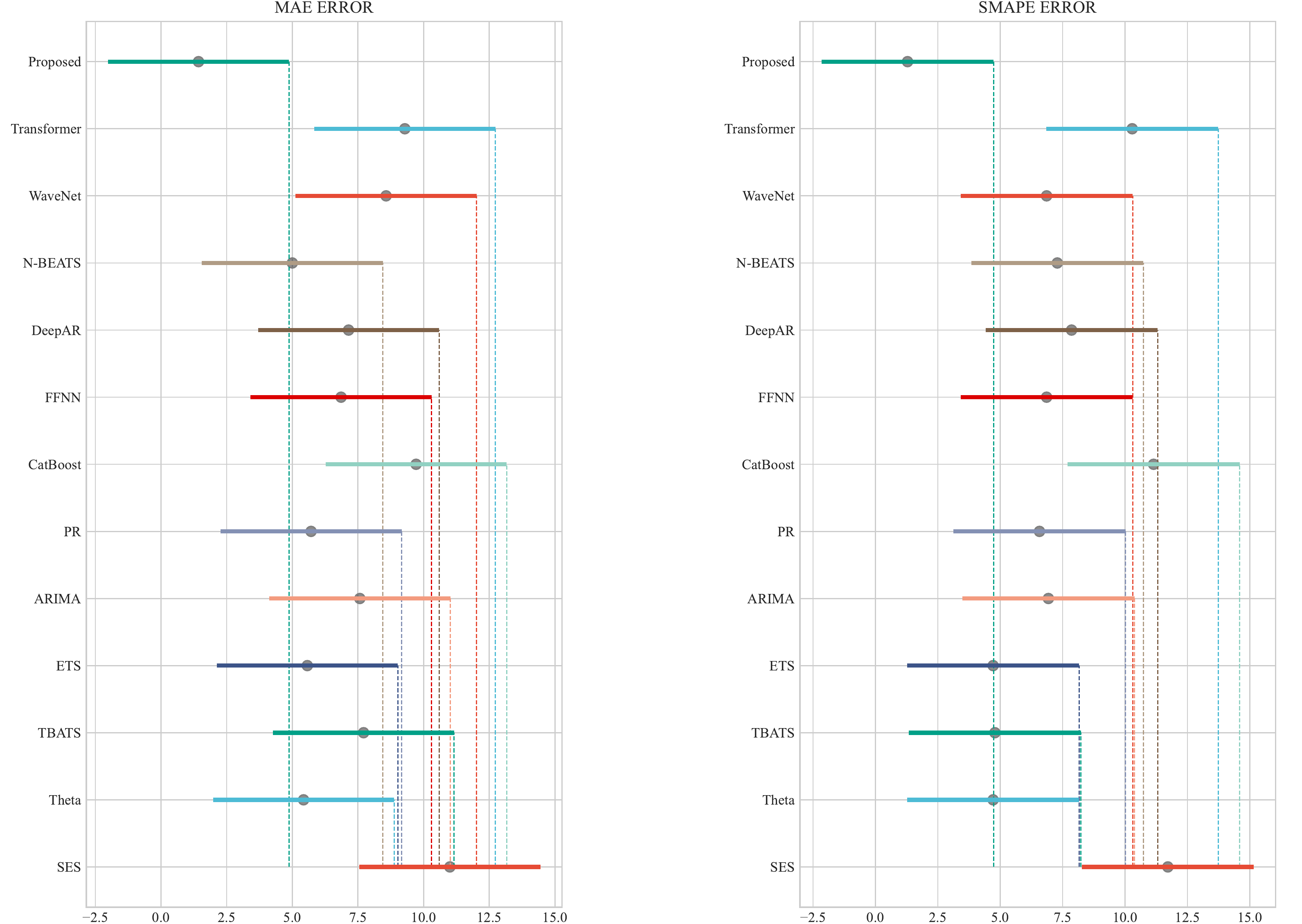}}
		\caption{Friedman test figure of proposed method and comparison methods by Nemenyi test on M1, M3 datasets. (The horizontal axis is the average order value, and the vertical axis is each algorithm. For each algorithm, a dot is used to display its average order value, and the horizontal line segment centered on the dot represents the size of the critical range.)}
	\end{figure}

	In order to compare the proposed method with the comparison method more intuitively, Friedman test and Nemenyi test \cite{pereira2015overview} are used to evaluate the model performance and the resulting Friedman test figure is shown in Fig.17. It can be seen that the proposed method outperforms the comparison method on M1 and M3 datasets.

	\section{Conclusion}
	
	Recently, time series forecasting in the framework under complex network is paid great attension. In this paper, a new time series forecasting method is presented.  The similarity generated after random walk is a key factor for forecasting in the network.  A new weighted node similairty is constructed. The results show the efficiency of the proposed method. In future work, the proposed method will continue to be improved.

	\section*{CRediT authorship contribution statement}
	\textbf{Tianxiang Zhan}: Conceptualization, Methodology, Software, Writing – original draft. \textbf{Fuyuan Xiao}: Writing – review \& editing, Project administration, Funding acquisition, Supervision.
	
	\section*{Declaration of competing interest}
	The authors declare that they have no known competing financial interests or personal relationships that could have appeared to influence the work reported in this paper.

	\section*{Data availability}
		Data will be made available on request.

	\section*{Acknowledgments}
		This research is supported by the National Natural Science Foundation of China (No. 62003280) and Chongqing Talents: Exceptional Young Talents Project (CQYC202105031).

	\bibliographystyle{elsarticle-num}
	\bibliography{References}
	
\end{document}